\documentclass[showpacs,preprint,amsmath,amssymb,nofootinbib,superscriptaddress]{revtex4}

\usepackage{graphicx}
\usepackage{dcolumn}
\usepackage{bm}
\usepackage{lscape}
\usepackage[usenames]{color}
\usepackage{ulem}

\usepackage{txfonts}

\begin{document}

\title{Global optical potential for nucleus-nucleus systems from 50 MeV/u to 400 MeV/u}

\author{T.~Furumoto}
\email{furumoto@yukawa.kyoto-u.ac.jp}
\affiliation{Yukawa Institute for Theoretical Physics, Kyoto University, Kyoto 606-8502, Japan}
\affiliation{RIKEN Nishina Center, Wako, Saitama 351-0198, Japan}

\author{W.~Horiuchi}
\affiliation{RIKEN Nishina Center, Wako, Saitama 351-0198, Japan}

\author{M.~Takashina}
\affiliation{Graduate School of Medicine, Osaka University, Suita, Osaka 565-0871, Japan}
\affiliation{Research Center for Nuclear Physics, Osaka University, Osaka 567-0047, Japan}

\author{Y.~Yamamoto}
\affiliation{RIKEN Nishina Center, Wako, Saitama 351-0198, Japan}

\author{Y.~Sakuragi}%
\affiliation{Department of Physics, Osaka City University, Osaka 558-8585, Japan}
\affiliation{RIKEN Nishina Center, Wako, Saitama 351-0198, Japan}

\date{\today}

\begin{abstract}
We present a new global optical potential (GOP) for nucleus-nucleus systems, including neutron-rich and proton-rich isotopes, in the energy range of  $50 \sim 400$ MeV/u.
The GOP is derived from the microscopic folding model with the complex $G$-matrix interaction CEG07 and the global density presented by S{\~ a}o Paulo group. 
The folding model well accounts for realistic complex optical potentials of nucleus-nucleus systems and reproduces the existing elastic scattering data for stable heavy-ion projectiles at incident energies above 50 MeV/u. 
We then calculate the folding-model potentials (FMPs) for projectiles of even-even isotopes, $^{8-22}$C, $^{12-24}$O, $^{16-38}$Ne, $^{20-40}$Mg, $^{22-48}$Si, $^{26-52}$S, $^{30-62}$Ar, and $^{34-70}$Ca, scattered by stable target nuclei of $^{12}$C, $^{16}$O, $^{28}$Si, $^{40}$Ca $^{58}$Ni, $^{90}$Zr, $^{120}$Sn, and $^{208}$Pb at the incident energy of 50, 60, 70, 80, 100, 120, 140, 160, 180, 200, 250, 300, 350, and 400 MeV/u. 
The calculated FMP is represented, with a sufficient accuracy, by a linear combination of 10-range Gaussian functions.
The expansion coefficients depend on the incident energy, the projectile and target mass numbers and the projectile atomic number, while the range parameters are taken to depend only on the projectile and target mass numbers.
The adequate mass region of the present GOP by the global density is inspected in comparison with FMP by realistic density.
The full set of the range parameters and the coefficients for all the projectile-target combinations at each incident energy are provided on a permanent open-access website together with a Fortran program for calculating the microscopic-basis GOP (MGOP) for a desired projectile nucleus by the spline interpolation over the incident energy and the target mass number. 
\end{abstract}


\maketitle

\section{Introduction}
The elastic scattering process contains valuable information about nuclear many-body dynamics in nuclear reactions induced by a nucleon or a composite-nucleus projectile on a target nucleus.
The elastic scattering observables (averaged over a certain range of incident energy) are known to be well described by a local or non-local one-body type complex potential called the optical potential.
The shape and the strength of the complex optical potential and their dependence on the incident energy and the mass and charge of the colliding system reflect, on one hand, the static properties of nuclear structure of the colliding system and the basic nucleon-nucleon ($NN$) interaction and, on the other hand, the effects of complicated non-elastic reaction processes that follow the elastic scattering process~\cite{FES58,HOD71,DNR,RAP82}. 

The optical potential is also indispensable to a reliable extraction of nuclear structure information from experimental data of various direct nuclear reactions such as inelastic scattering and transfer reactions.
Namely, most direct reaction theories, such as the distorted-wave Born approximation (DWBA), distorted-wave impulse approximation (DWIA) and coupled-channels (CC) method, incorporate the optical potential as the distorting potentials in the entrance or exit channels in the analyses of experimental data~\cite{DNR}.
It is often the case that the calculated results strongly depend on the choice of the distorting potential.
Therefore, it is very important to establish reliable optical potentials (or more generally interaction models between colliding nuclear systems) to survey unknown properties of nuclear structure and reaction dynamics through the experimental data.

Historically, phenomenological optical potentials (POPs) were deduced by analyzing individual elastic scattering data for various targets and incident energies. The systematic analyses of a large body of scattering data enabled us to find systematic behavior of potential parameters with the variation of the mass and charge of the target nucleus as well as the incident energy, which led to a proposal of the global optical potential (GOP).
The GOP for nucleon-nucleus scattering has a long history for more than four decades~\cite{WAT69,BG69,NAD81} and nowadays it is in almost established stage covering whole stable target nuclei and incident energies from nearly zero to 1 GeV~\cite{VAR91,HAM90,COO93,KON03,HAN10}.
For deuteron-nucleus scattering, various GOPs have been proposed~\cite{PP63,DAE80,BOJ88,AN06,HAN06} that are also applicable to a wide range of incident energy and target mass number.

The study of optical potentials for composite projectiles heavier than deuteron is still in a developing stage and far from satisfactory level, partly because of insufficient number of experimental data on elastic scattering compared with nucleon and deuteron cases and, more essentially, because of the composite nature of the system which makes the nuclear reaction dynamics more complicated, such as mutual excitations, breakup, transfer, fusion etc. The increase of the number of reaction channels leads to a strong absorption of flux from the elastic-scattering channel, which makes it difficult to uniquely determine the shape and strength of the optical potential~\cite{DNR}.

Nevertheless, a number of efforts have been made to deduce the optical potentials for composite projectiles heavier than deuteron and GOPs were also proposed for some light ions, such as $^{3}$He~\cite{LI07,LIA09}, $\alpha$ particles~\cite{NOL87,AVR94,KUM06} and some light heavy-ion projectiles $^{6,7}$Li~\cite{COO82} and $^{6}$He~\cite{KUC09}. However, the range of the incident energy and target mass number covered by those GOPs is very limited compared with those for nucleon and deuteron scattering. 

For heavy-ion projectiles, a number of elastic-scattering experiments have been done and optical potentials have been deduced where ambiguity and uniqueness of the deduced optical potentials were discussed and some sort of systematics were studied~\cite{BRAN97}.
However, the number of experimental data is far from satisfactory level to extract GOPs, except for some limited scattering systems for which the proposed ``GOPs'' are only applicable to a limited energy range~\cite{CHA02,KOB84}. 

Nowadays the experimental facilities have been developed intensively. 
The radioactive isotope beam is produced with a considerable intensity up to the driplines for light elements and its energy reaches up to 400 MeV/u. 
In order to deduce meaningful information on the properties and nuclear structure of such rare isotopes through reaction experiments, it is essentially important to establish the optical potentials, or any alternative interaction models, for those unstable heavy-ion systems up to the highest energy region, say 400 MeV/u. 
In most radioactive ion beam experiments, however, it is very difficult to measure reliable elastic-scattering cross sections, particularly to obtain a precise angular distribution necessary for extracting reliable optical potentials. 
Therefore, one needs a reliable GOP or any alternative theoretical interaction models.

The global description of the optical potential has been attempted in the view point of the microscopic interaction models. 
For a nucleon-nucleus system, the most successful description of the complex optical potentials and the elastic scattering data is obtained by the single-folding model approach with complex $G$-matrix interactions~\cite{BRI77, ROO77, BRI78, RIK841, RIK842, JLM77, JLM98, CEG83, CEG86, MEL00, CEG07, MIN10} based on the Br\"{u}ckner Hartree-Fock theory.
The energy and target-mass dependences of the geometrical shape and strength of the potential are explained by various kinds of medium effects of effective interaction ($G$-matrix) in nuclear medium as well as the exchange terms and the non-locality, in addition to the genuine energy dependence of the free-space $NN$ interaction used and the nucleon density distribution of the target nucleus.

Microscopic description of the nucleus-nucleus interaction has also been studied for decades on the basis of double-folding model (DFM) with the $G$-matrix interactions. 
The most successful models in the earlier stage are those based on the M3Y interaction~\cite{M3Y77,SAT79} or its extension having various types of density dependence~\cite{KOB82, KOB84, FARID85, KHO94, BRAN97, KHO97}.
However, all these interaction models for heavy-ion optical potential predict only the real part of the potential and the phenomenological imaginary part has to be added by hand. 
The JLM interaction~\cite{JLM77} had been the only complex $G$-matrix interaction applied to the construction of complex heavy-ion optical potential~\cite{CAR96, TRA00, BLA05, FUR06}. 
The JLM interaction is known to be successful for nucleon-nucleus system but it has a crucial problem in the application to the nucleus-nucleus systems. 
Namely, the calculated DFM potential with the JLM interaction is too strong both in the real and imaginary parts compared to the POP evaluated from experimental data and needs a strong renormalization.
The origin of the renormalization problem in the DFM with JLM interaction is found to be originated from improper prescriptions (average prescriptions) for evaluating the local density in the DFM calculation~\cite{FUR06}, which was inevitable for the JLM
interaction that was given only for nuclear matter with density below the saturation density $\rho_0$.

In order to overcome this problem, three of the present authors have recently constructed a new complex $G$-matrix interaction called CEG07~\cite{CEG07, FUR09}, that is applicable for nucleon density up to twice the saturation density, $\rho \approx 2\rho_0$ in the energy range of $E = 20 \sim 400$ MeV/u and they have proposed the DFM with the CEG07 interaction that gives a reliable complex optical potential for nucleus-nucleus systems~\cite{FUR09R, FUR09, FUR10}. 
It is shown that the DFM with the new $G$-matrix predicts the proper strength and the shape of the nucleus-nucleus optical potentials and reproduces almost all the existing experimental data for elastic scattering of $^{12}$C and $^{16}$O projectiles by various stable target nuclei in the energy-range of $E = 70 \sim 135$ MeV/u~\cite{FUR09R, FUR09} with essentially no renormalization.
It is interesting to note that the folding model also predicts that the real part of heavy-ion optical potentials changes its sign from negative (attractive) to positive (repulsive) around $E = 200 \sim 300$ MeV/u~\cite{FUR10}.
In addition to the elastic scattering, the CEG07 folding model has also been applied to the coupled-channel analyses of inelastic scattering with considerable success~\cite{TAK10}.
The present DFM with CEG07 is now the most reliable microscopic interaction model at present for constructing the complex optical potential for nucleus-nucleus scattering systems. 

However, one needs rather complicated and elaborated numerical tasks to calculate the DFM potential with CEG07 because of its precise treatment of density dependence as well as that of the finite-range exchange component and, hence, it will not be a better choice to perform heavy numerical calculations to analyze individual scattering experiments for each combination of projectile-target system at each scattering energy. It would rather be desirable to construct reliable GOP on the basis of the reliable microscopic interaction model. 
\begin{figure*}[t]
\begin{center}
\includegraphics[width=16cm]{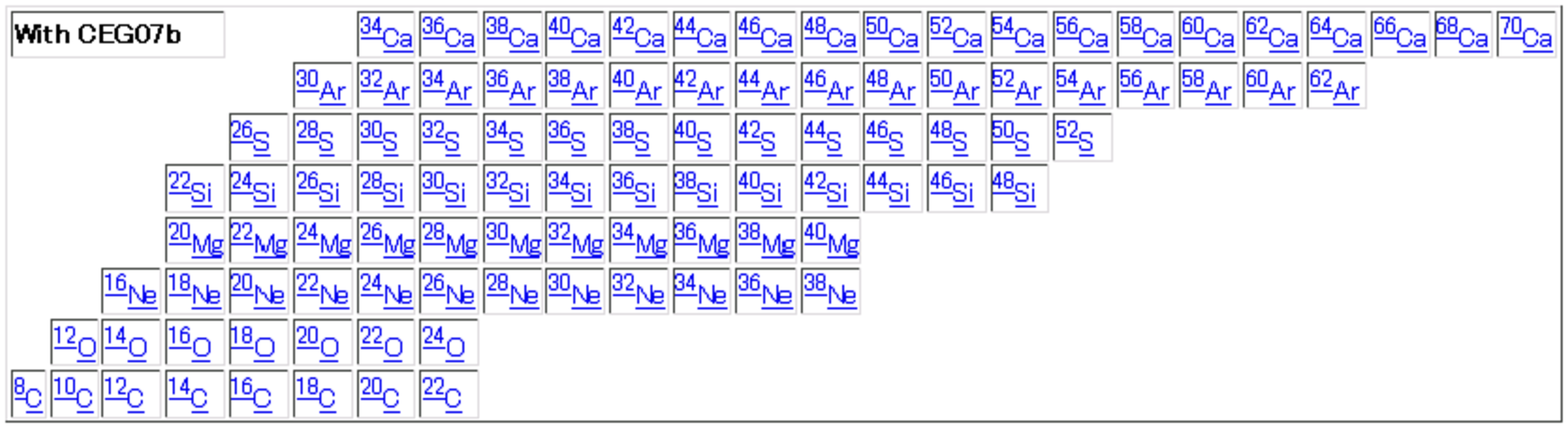}
\caption{\label{fig01} (Color online) The table on the web site~\cite{furumoto}.}
\end{center}
\end{figure*}

In the present paper, we apply the present microscopic folding model with the CEG07 interaction to the construction of GOP for scattering of heavy ions including neutron-rich and proton-rich unstable isotopes over the wide range of nuclear chart.
To this end, we first calculate the DFM potentials for projectiles of even-even isotopes, $^{8-22}$C, $^{12-24}$O, $^{16-38}$Ne, $^{20-40}$Mg, $^{22-48}$Si, $^{26-52}$S, $^{30-62}$Ar, and $^{34-70}$Ca (see Fig.~\ref{fig01}), scattered by stable target nuclei of $^{12}$C, $^{16}$O, $^{28}$Si, $^{40}$Ca $^{58}$Ni, $^{90}$Zr, $^{120}$Sn, and $^{208}$Pb at certain steps of incident energy, 
say $E$ = 50, 60, 70, 80, 100, 120, 140, 160, 180, 200, 250, 300, 350, and 400 MeV/u. 
The calculated folding potential is represented, with a sufficient accuracy, by a linear combination of 10-range Gaussian functions.
The expansion coefficients depend on the incident energy, the projectile and target mass numbers and the projectile atomic number, while the Gauss-range parameters are taken to depend only on the projectile and target mass numbers.
The full set of the range parameters and the coefficients for all the projectile-target combinations at each incident energy are then stored on a permanent open-access website~\cite{furumoto} together with a Fortran program for calculating the microscopic-basis GOP (MGOP) for a requested projectile nucleus by the spline interpolation of the parameter sets over the incident energy and the target mass number. 
These parameter sets and the program will be provided to open use for a wide range of nuclear reaction analyses.

In the next section, we briefly describe the theoretical frame of the microscopic folding model with the CEG07 interaction together with the parameterizations of the calculated DFM potentials in a form of GOP, where we discuss the global parametrization of the input nucleon density distributions. 
In sect.III, we first compare elastic scattering cross sections calculated with the  microscopic DFM potential as well as its equivalent MGOP with the existing experimental data for stable-nucleus heavy-ion projectiles, $^{12}$C, $^{16}$O and $^{40}$Ar, by various target nuclei at various incident energy, to confirm the applicability of the present MGOPs. 
We them calculate elastic-scattering cross sections with the MGOPs for projectile of unstable nuclei including those near the driplines and discuss the isotope dependence of the predicted cross sections that reflect different density distributions through the DFM procedure.
The final section will be devoted to conclusion.

\section{Global Optical Potential}
\subsection{Double folding model}
We construct the nucleus-nucleus optical model potential based on the DFM using the complex $G$-matrix interaction CEG07. 
In the previous work~\cite{FUR09R,FUR09}, it was found that the three-body force effect, particularly its repulsive component, plays a crucial role to obtain the proper strength and shape of the nucleus-nucleus potential consistent with the experimental data.
Thus, we use the CEG07b version of the CEG07 interaction that includes the three-body force effect.

The microscopic nucleus-nucleus potential can be written as a Hartree-Fock type potential: 
\begin{eqnarray}
U_{\rm{F}}&=&\sum_{i\in A_{1},j\in A_{2}}{[<ij|v_{\rm{D}}|ij>+<ij|v_{\rm{EX}}|ji>]}\\
&=&U_{\rm{D}}+U_{\rm{EX}},
\end{eqnarray}
where $v_{\rm{D}}$ and $v_{\rm{EX}}$ are the direct and exchange parts of complex $G$-matrix interaction. 
The exchange part is a nonlocal potential in general but can be treated as a local potential by using the plane-wave representation for the $NN$ relative motion \cite{SIN75, SIN79}. 
The direct and exchange parts of the localized potential are then written in the standard form of the DFM potential as 
\begin{eqnarray}
U_{\rm{D}}(R)&=&\int{\rho_{1}(\bm{r}_1) \rho_{2}(\bm{r}_2) v_{\rm{D}}(s; \rho, E)d\bm{r}_1 d\bm{r}_2} \\
&=&\int \Bigl\{ \rho^{\rm{(p)}}_{1}(\bm{r}_1) \rho^{\rm{(p)}}_{2}(\bm{r}_2) v^{\rm{(pp)}}_{\rm{D}}(s; \rho, E) \nonumber \\
&&\ \ \ \ +\rho^{\rm{(p)}}_{1}(\bm{r}_1) \rho^{\rm{(n)}}_{2}(\bm{r}_2) v^{\rm{(pn)}}_{\rm{D}}(s; \rho, E) \nonumber \\
&&\ \ \ \ +\rho^{\rm{(n)}}_{1}(\bm{r}_1) \rho^{\rm{(p)}}_{2}(\bm{r}_2) v^{\rm{(np)}}_{\rm{D}}(s; \rho, E) \nonumber \\
&&\ \ \ \ +\rho^{\rm{(n)}}_{1}(\bm{r}_1) \rho^{\rm{(n)}}_{2}(\bm{r}_2) v^{\rm{(nn)}}_{\rm{D}}(s; \rho, E) \Bigr\} d\bm{r}_1 d\bm{r}_2,
\end{eqnarray}
where $\bm{s}=\bm{r}_2-\bm{r}_1+\bm{R}$, and
\begin{eqnarray}
U_{\rm{EX}}(R)&=&\int{\rho_{1}(\bm{r}_1, \bm{r}_1+\bm{s}) \rho_{2}(\bm{r}_2, \bm{r}_2-\bm{s}) v_{\rm{EX}}(s; \rho, E)} \nonumber \\
&&\ \ \ \ \times \exp{ \left[ \frac{i\bm{k}(R)\cdot \bm{s}}{M} \right] } d\bm{r}_1 d\bm{r}_2 \\ 
&=&\int \Bigl\{ \rho^{\rm{(p)}}_{1}(\bm{r}_1, \bm{r}_1+\bm{s}) \rho^{\rm{(p)}}_{2}(\bm{r}_2, \bm{r}_2-\bm{s}) v^{\rm{(pp)}}_{\rm{EX}}(s; \rho, E) \nonumber \\
&&\ \ \ \ +\rho^{\rm{(p)}}_{1}(\bm{r}_1, \bm{r}_1+\bm{s}) \rho^{\rm{(n)}}_{2}(\bm{r}_2, \bm{r}_2-\bm{s}) v^{\rm{(pn)}}_{\rm{EX}}(s; \rho, E) \nonumber \\
&&\ \ \ \ +\rho^{\rm{(n)}}_{1}(\bm{r}_1, \bm{r}_1+\bm{s}) \rho^{\rm{(p)}}_{2}(\bm{r}_2, \bm{r}_2-\bm{s}) v^{\rm{(np)}}_{\rm{EX}}(s; \rho, E) \nonumber \\
&&\ \ \ \  +\rho^{\rm{(n)}}_{1}(\bm{r}_1, \bm{r}_1+\bm{s}) \rho^{\rm{(n)}}_{2}(\bm{r}_2, \bm{r}_2-\bm{s}) v^{\rm{(nn)}}_{\rm{EX}}(s; \rho, E) \Bigr\} \nonumber \\
&&\ \ \ \ \times \exp{ \left[ \frac{i\bm{k}(R)\cdot \bm{s}}{M} \right] } d\bm{r}_1 d\bm{r}_2, \label{eq:ex}
\end{eqnarray}
where $E$ is the incident energy per nucleon (MeV/u).
The superscript of $\rho$ (p and n) indicates the proton and neutron densities, respectively. 
The direct (D) and exchange (EX) parts of the proton-proton (pp), proton-neutron (pn), neutron-proton (np), and neutron-neutron (nn) $G$-matrix interactions are written as
\begin{eqnarray}
v_{\rm{D, EX}}^{(\rm{pp, nn})}&=&\frac{1}{4}(v^{01}\pm 3v^{11}),
\label{eq:coption11}
\\
v_{\rm{D, EX}}^{(\rm{pn, np})}&=&\frac{1}{8}(\pm v^{00}+v^{01}+3v^{10}\pm 3v^{11}),
\label{eq:coption12}
\end{eqnarray}
in terms of $v^{ST}$, the spin-isospin component ($S$=0 or 1 and $T$=0 or 1) of the $G$-matrix interaction.
Here, the upper and lower part of the double-sign symbols ($\pm$) in the r.h.s. correspond to the direct (D) and exchange (EX) parts, respectively.
Here, the exponential function in Eq.~(\ref{eq:ex}) is transformed into the spherical Bessel function, $j_{0} (\frac{k(R)s}{M})$, by the multipole expansion.
The $k(R)$ is the local momentum for nucleus-nucleus relative motion, the magnitude of which is given by 
\begin{equation}
k^2(R)=\frac{2mM}{\hbar^2}[E_{\rm{c.m.}}-{\rm{Re}}U_{\rm{F}}(R)-V_{\rm{coul}}(R)], \label{eq:kkk}
\end{equation}
where $M=A_1A_2/(A_1+A_2)$, $E_{\rm{c.m.}}$ is the center-of-mass energy, $m$ is the nucleon mass and {\it $V_{\rm{coul}}$} is the Coulomb potential. 
Here, the Coulomb potential {\it $V_{\rm{{coul}}}$} is also obtained by folding the $NN$ Coulomb potential with the proton density distributions of the projectile and target nuclei. 
$A_{1}$ and $A_{2}$ are the mass numbers of the projectile and target, respectively. 
The exchange part is calculated self-consistently on the basis of the local energy approximation through Eq.~(\ref{eq:kkk}). 
The density matrix $\rho(\bm{r}, \bm{r}')$ is approximated in the same
manner as in Ref.~\cite{NEG72}; 
\begin{equation} 
\rho (\bm{r}, \bm{r}')
=\frac{3}{k^{\rm{eff}}_{\rm{F}}s}j_{1}(k^{\rm{eff}}_{\rm{F}}s)\rho \Big(\frac{\bm{r}+\bm{r}'}{2}\Big), 
\label{eq:exchden}
\end{equation}
where $k^{\rm{eff}}_{\rm{F}}$ is the effective Fermi momentum \cite{CAM78} defined by
\begin{equation}
k^{\rm{eff}}_{\rm{F}} 
=\Big( (3\pi^2 \rho )^{2/3}+\frac{5C_{\rm{s}}[\nabla\rho]^2}{3\rho^2}
+\frac{5\nabla ^2\rho}{36\rho} \Big)^{1/2}, \;\; 
\label{eq:kf}
\end{equation}
where we adopt $C_{\rm{s}} = 1/4$ following Ref.~\cite{KHO01}. 
In the present calculations, we employ the so-called frozen density approximation (FDA) for evaluating the local density.
In the FDA, the density-dependent $NN$ interaction is assumed to feel the local density defined as the sum of densities in colliding nuclei evaluated at the mid-point of the interacting nucleon pair:  
\begin{equation}
\rho = \rho_{1} + \rho_{2} \label{eq:fda}
\end{equation}
where the local densities are evaluated at the position of each nucleon for the direct part and at the middle point of the interacting nucleon pair for the exchange part~\cite{KHO00,KAT02}. 
The FDA has been widely used also in the standard DFM calculations~\cite{KHO94, KHO97, KHO01, KAT02, SAT79, FUR09}.
In Ref.~\cite{FUR09} it is confirmed that FDA is the best prescription in the case with CEG07b to reproduce the data. 

\subsection{Input density}
The input density is important to construct the global optical potential in the folding procedure. 
In this paper, we adopt the S{\~ a}o Paulo density provided by S{\~a}o Paulo group~\cite{CHA02} as a global density. 
The S{\~a}o Paulo density has the functional form of the two-parameter Fermi, as
\begin{equation}
\rho_{\rm{p, n}} (r) = \frac{\rho_{0}}{1+\exp(\frac{r - R_{\rm{p, n}}}{a_{\rm{p, n}}})}, 
\end{equation}
where, the subscript of $\rho$, $R$, and $a$ (p and n) indicates the proton and neutron parts, respectively. 
The parameters of $R_{\rm{p, n}}$ and $a_{\rm{p, n}}$ are obtained in Ref.~\cite{CHA02}, as
\begin{eqnarray}
R_{\rm{p}}&=&1.81Z^{1/3}-1.12, \\
R_{\rm{n}}&=&1.49N^{1/3}-0.79, \\
a_{\rm{p}}&=&0.47-0.00083Z, \\
a_{\rm{n}}&=&0.47+0.00046N,
\end{eqnarray}
where, $Z$ and $N$ are the proton and neutron number, respectively. 

Here, we note that the S{\~a}o Paulo density is fitted to the calculated results of the Dirac-Hartree-Bogoliubov model in the stable nuclear region. 
It is nontrivial that the S{\~a}o Paulo density is useful in the unstable nuclear region.
However, our purpose is the construction of global optical potential with various projectiles including the proton-rich or neutron-rich nucleus. 
Therefore, we compare the results obtained by the S{\~a}o Paulo density with those by a theoretical framework aimed at precise description of nucleon density for unstable nucleus region up to driplines.
As shown in Sec.~III. B, the GOP with the S{\~a}o Paulo density gives the similar results to the results by the FMP with such realistic density, except in the extreme case, such as near or on the dripline.
For such realistic densities, we adopt those proposed by Niigata group for Carbon isotopes~\cite{HOR06-1,HOR06-2,HOR07}, for Oxygen isotopes~\cite{ABU09}, and for $^{30}$Ne nucleus \cite{HOR10}, and we call them as Niigata density. 
Niigata densities have already been applied to the analyses of total reaction cross sections in the Glauber model and found to well reproduce the experimental data~\cite{HOR07,ABU09}. 
The Niigata density is obtained by a Slater determinant based on empirical nucleon separation energies. 
A core + two-nucleon model is employed for $^{16}$C~\cite{HOR06-1} and $^{22}$C~\cite{HOR06-2} in order to take into account some dynamical effects beyond the simple mean field model. 
Since $^{21}$C is unstable with respect to a neutron emission, it is more realistic to describe $^{22}$C with the three-body model. 

\begin{figure}[thb]
\begin{center}
\includegraphics[width=8cm]{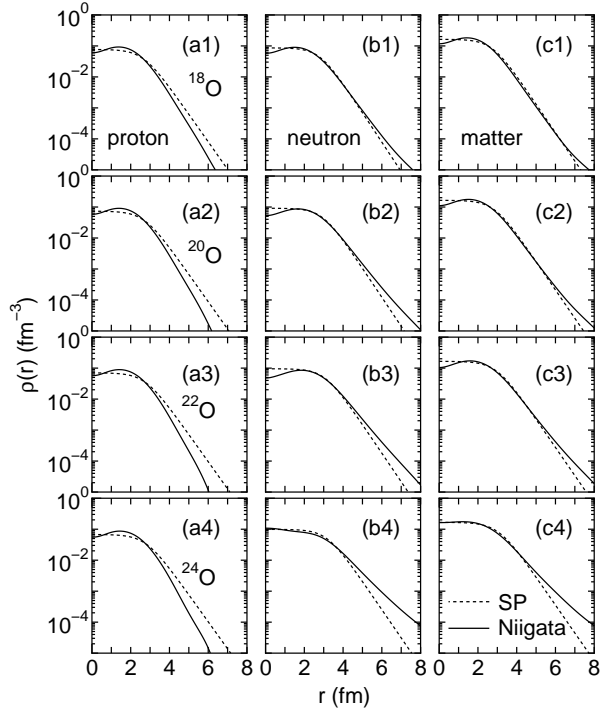}
\caption{\label{fig02} Comparison the S{\~a}o Paulo (SP) density with the Niigata one for the Oxygen isotopes. 
The solid and dotted curves are the Niigata and S{\~a}o Paulo (SP) densities, respectively.
The alphabets of the subscription (a, b, and c) indicate the proton, neutron, and matter densities, respectively.
The numbers of the subscription (1, 2, 3, and 4) indicate the mass number ($A =$ 18, 20, 22, and 24) for the Oxygen isotopes, respectively.
}
\end{center} 
\end{figure}
Figure~\ref{fig02} shows the proton, neutron, and matter density distributions of the S{\~a}o Paulo and Niigata densities for the Oxygen isotopes.
The two kinds of density distributions have similar form near the stable region, especially for the matter distribution.
On the other hand, the difference is clearly seen on the dripline.
The large difference in the middle and long range parts of the matter density distribution reflects the difference in the neutron density distribution. 
In the next section, we compare the S{\~a}o Paulo density with the Niigata one in the calculated elastic cross section and investigate the adequate mass region of the present GOP with the S{\~a}o Paulo density.

\subsection{Functional form of global optical potential}
We parameterize the nuclear part of the DFM potential in a functional form.
The real and imaginary parts of the calculated DFM potential are represented 
in terms of a linear combination of 10-range Gaussian form, respectively, with sufficient accuracy as
\begin{eqnarray}
V_{\rm{F}}(R)\cong \sum_{n=1}^{10}{\left\{ \alpha_{n}\exp{\left( -\frac{R^{2}}{\gamma_{n}^{2}}\right) }\right\} }\; \equiv V_{\rm{GOP}}(R), \\ 
W_{\rm{F}}(R)\cong \sum_{n=1}^{10}{\left\{ \beta_{n}\exp{\left( -\frac{R^{2}}{\gamma_{n}^{2}}\right) }\right\} }\; \equiv W_{\rm{GOP}}(R), 
\end{eqnarray}
where, $V_{\rm{F}}(R)$ and $W_{\rm{F}}(R)$ are the real and imaginary parts of the FMP.
The $V_{\rm{GOP}}(R)$ and $W_{\rm{GOP}}(R)$ are defined as the real and imaginary parts of the GOP, and, 
\begin{eqnarray}
\alpha_{n}&=&\alpha_{n}(A_{p}, Z_{p}, A_{t}, E)\; ,\\
\beta_{n}&=&\beta_{n}(A_{p}, Z_{p}, A_{t}, E)\; ,\\
\gamma_{n}&=&0.45\left( \frac{n+8}{18}\right) (A_{p}^{1/3} + A_{t}^{1/3} + 1)\; . \label{eq:fit}
\end{eqnarray}
Here, $A_{p}$, $Z_{p}$, $A_{t}$, and $E$ are the mass number of the projectile, the proton number of the projectile, the mass number of the target, and the incident energy per nucleon, respectively. 
The ``$+1$'' in Eq.~(\ref{eq:fit}) is introduced to reflect the finite range of the $NN$ interaction.
The factor $(\frac{n+8}{18})$ with $n=1\sim 10$ is taken to be an arithmetic progression from 0.5 to 1.0. 

We calculate the potentials for the incident even-even nuclei ($^{8-22}$C, $^{12-24}$O, $^{16-38}$Ne, $^{20-40}$Mg, $^{22-48}$Si, $^{26-52}$S, $^{30-62}$Ar, and $^{34-70}$Ca) by the $^{12}$C, $^{16}$O, $^{28}$Si, $^{40}$Ca $^{58}$Ni, $^{90}$Zr, $^{120}$Sn, and $^{208}$Pb targets at 
50, 60, 70, 80, 100, 120, 140, 160, 180, 200, 250, 300, 350, and 400 MeV/u.
Here we only consider a stable nucleus as a target. 
The desired optical potential is obtained by the spline interpolation 
over the incident energy and the target mass number.
The parameters, $\alpha$ and $\beta$, are available in Ref.~\cite{furumoto}. 
The total number of the parameters provided is about 280,000. 
For a convenient use we also provide the program source "MGOP" in Ref.~\cite{furumoto} which includes functions to generate the parameters and construct the optical potential.

\begin{figure}[thb]
\begin{center}
\includegraphics[width=8cm]{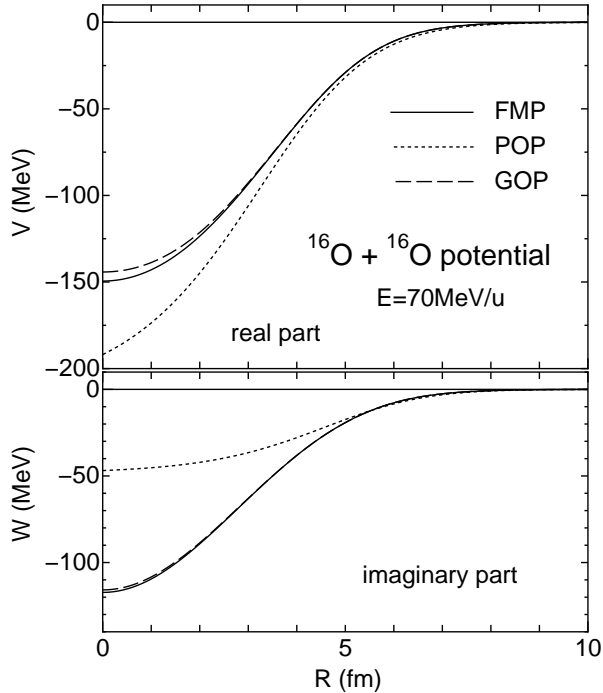}
\caption{\label{figXX} Comparison GOP with FMP and phenomenological optical potential (POP) for the $^{16}$O + $^{16}$O system at $E =$ 70 MeV/u. 
The solid, dotted, and dashed curves are the FMP, POP, and GOP, respectively.
The POP is taken from Ref.~\cite{KHOA00}}
\end{center} 
\end{figure}
Figure~\ref{figXX} shows the calculated FMP, phenomenological optical potential (POP), and GOP for the $^{16}$O + $^{16}$O system at $E =$ 70 MeV/u.
The GOP well fit to the FMP by a linear combination of 10-range Gaussian form, except for the most inner part of the real part ($R =$ 0-2 fm).
The real parts of FMP and GOP are slightly shallow in the comparison with that of POP.
On the other hand, the FMP and GOP give the large imaginary potential.
However, the both of the tail part almost becomes same strength.
We will discuss this difference and the ambiguity of the imaginary part for the elastic cross section in the next section. 

In this paper, we do not provide the Coulomb part of GOP in a functional form and the standard Coulomb potential of a uniform charge is supposed to be used in the application of the nuclear part of GOPs proposed here.  
We use the radius of uniform charge, $R_{\rm{C}} = 1.3 (A_{p}^{1/3} + A_{t}^{1/3})$ fm.

\section{Results}
\subsection{Comparison with experimental data}
We now calculate the elastic scattering cross section with the folding model potential (FMP) and the present GOP fitted by 10-range Gaussian form. 
The results by these potentials are compared with the experimental data. 
The CEG07 folding model has only one parameter $N_{\rm{W}}$ \cite{FUR09} and the optical potential is given by
\begin{equation}
U_{\rm opt}(R) = V_{\rm F}(R) + i N_{\rm{W}} W_{\rm F}(R).
\end{equation}
In this paper, we also discuss the energy and target dependences of this $N_{W}$ value. 

\begin{figure}[tbh]
\begin{center}
\includegraphics[width=7cm]{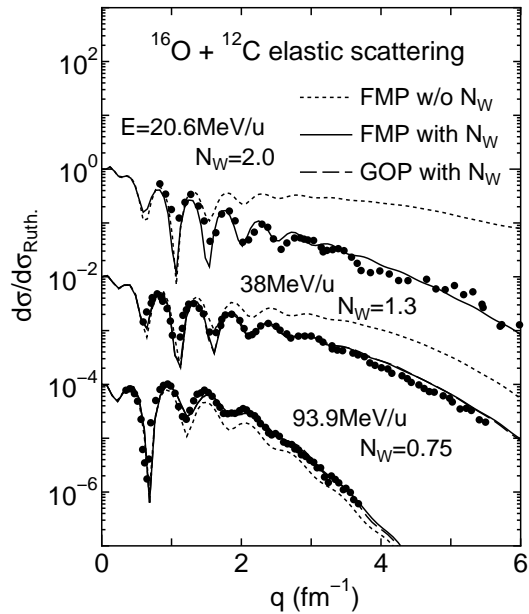}
\caption{\label{fig03} Elastic scattering for the $^{16}$O + $^{12}$C system at various energies.
Abscissa $q$ is the momentum transfer. 
The solid and dotted curves are the results by the FMP with the best fit and constant values ($N_{W}$ = 1.0), respectively. 
The dashed curves are the results by the present GOP with the $N_{W}$ value. 
The experimental data is taken from Refs.~\cite{DEM04,BRA86,ROU88}.}
\end{center} 
\end{figure}

\begin{figure}[tbh]
\begin{center}
\includegraphics[width=7cm]{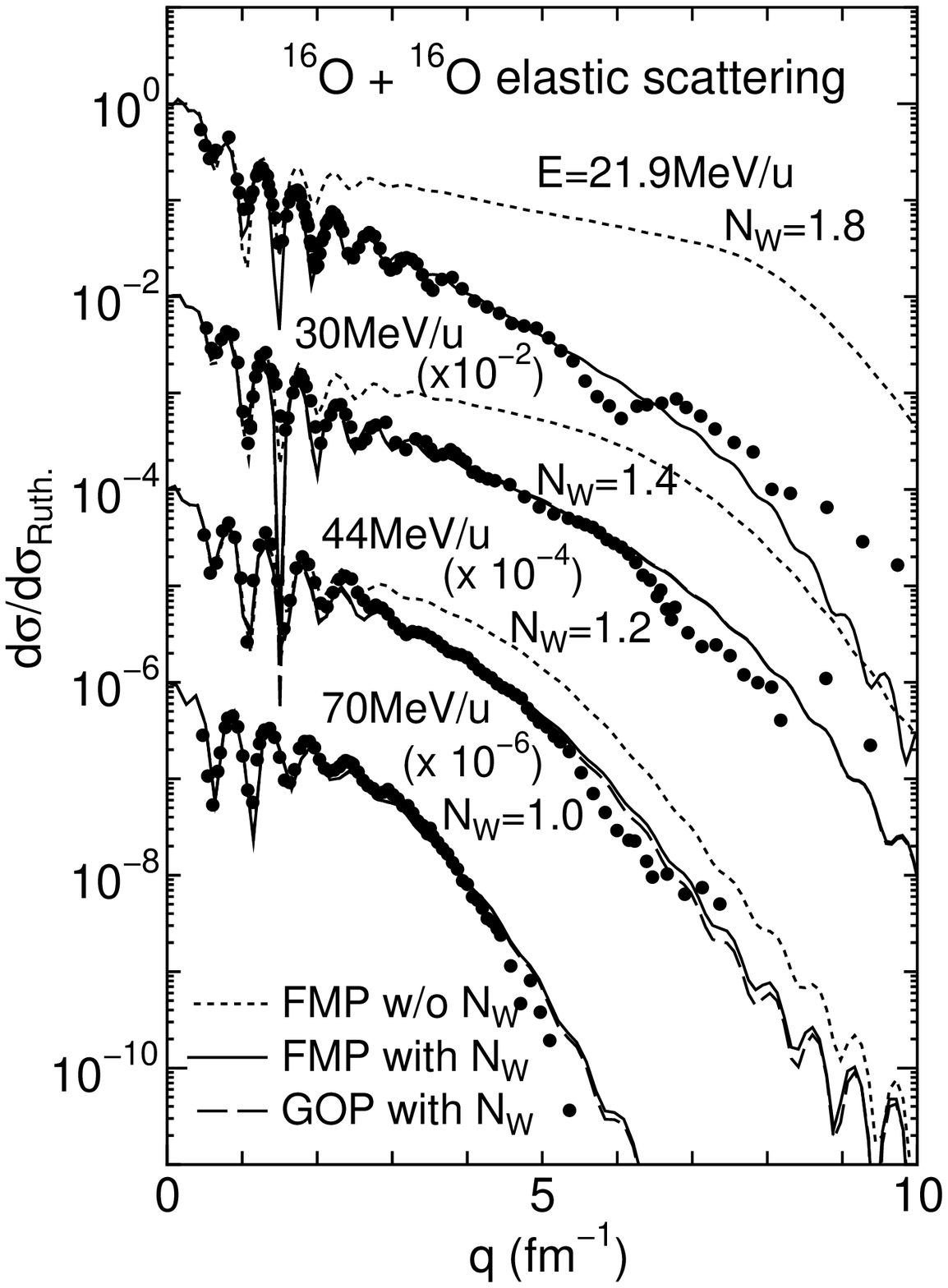}
\caption{\label{fig04} Same as Fig.~\ref{fig03} but for the $^{16}$O + $^{16}$O system. 
The experimental data is taken from Refs.~\cite{KHOA95,BAR96,NUO98,KHOA00}.}
\end{center} 
\end{figure}

\begin{figure}[tbh]
\begin{center}
\includegraphics[width=7cm]{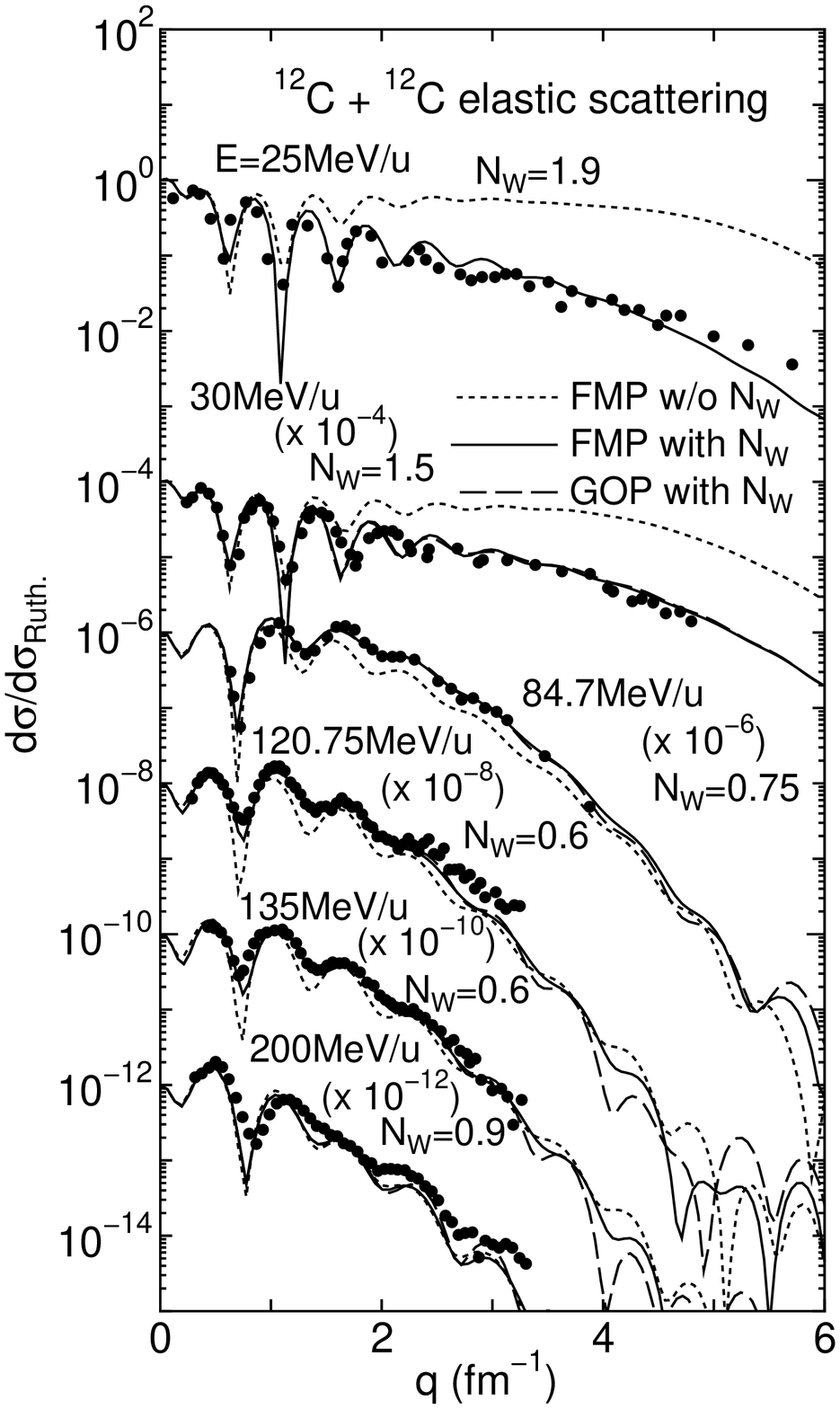}
\caption{\label{fig05} Same as Fig.~\ref{fig03} but for the $^{12}$C + $^{12}$C system. 
The experimental data is taken from Refs.~\cite{BOH82,BUE84,BUE81,ICH94,HOS87}.}
\end{center} 
\end{figure}

We first test the FMP in the elastic scattering of some stable nuclei and investigate 
how the present FMP reproduces the experimental data free of renormalization factor.
It should be emphasize that we do not renormalize the real part of FMP in all the calculations shown below and we investigate the need of the
renormalization to the imaginary part of the original FMP.

Figures~\ref{fig03}, \ref{fig04} and \ref{fig05} show elastic scattering cross section for the $^{16}$O + $^{12}$C, $^{16}$O + $^{16}$O and 
$^{12}$C + $^{12}$C systems at various energies. 
The solid and dotted curves are the results by the FMP with and without the renormalization factor for the imaginary part ($N_{W}$), respectively. 
The calculated cross sections reasonably reproduce the experimental data without the modification of $N_{W}$ from unity, except for the low energy cases below 50 MeV/u.
We thus conclude that the present FMP with CEG07b gives a reasonable account of elastic scattering cross sections at the incident energy above 50 MeV/u in free of any adjustable parameters, although more perfect fit can be attained by a slight adjustment of $N_{W}$.

The dashed curves are the results by the present GOP with the same $N_{W}$ values. 
The results by the present GOP at 20.6 MeV/u for $^{16}$O+$^{12}$C, 21.9 MeV/u for $^{16}$O+$^{16}$O and 25 MeV/u for $^{12}$C + $^{12}$C are not shown in these figures because we prepare the GOP in the energy range of 30 $\sim$ 400 MeV/u. 
The FMP and the present GOP give almost the same cross section, which implies that the fitting accuracy is enough to describe the elastic scattering given by the original FMP, although minor deviations from the original results  are seen in high $q$ region ($q\geq 4$ fm$^{-1}$) of the $^{12}$C + $^{12}$C scattering at higher incident energies above 100 MeV/u.
The calculated elastic cross sections with GOP are close to the experimental data even without $N_{W}$ above 50 MeV/u and we decide to provide the GOP parameters for the incident energies above 50 MeV/u in the present paper.
Here, we note that the FMP and GOP well reproduce the data despite the large difference of the imaginary potential seen in Fig.~\ref{figXX}.
This implies that the experimental data cannot probe the imaginary-potential strength at short distance. 

\begin{figure}[tbh]
\begin{center}
\includegraphics[width=7cm]{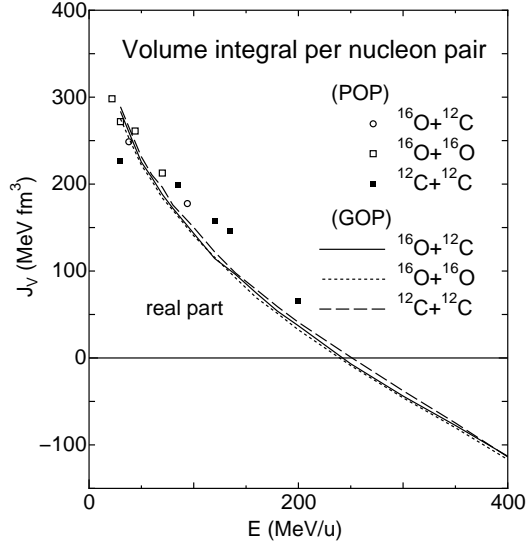}
\caption{\label{figYY} Volume integral per nucleon pair of the GOP and POP for the $^{16}$O + $^{12}$C, $^{16}$O + $^{16}$O, and $^{12}$C + $^{12}$C systems.
POPs are taken from Refs.~\cite{DEM04,BRA86,ROU88,KHOA00,BOH82,BUE84,BUE81,ICH94,HOS87}.}
\end{center} 
\end{figure}
Figure~\ref{figYY} shows the real part of the volume integral per nucleon pair of the GOP and POP for the $^{16}$O + $^{12}$C, $^{16}$O + $^{16}$O, and $^{12}$C + $^{12}$C systems.
The volume integral per nucleon pair is defined as
\begin{equation}
J_{V} = - 4 \pi \int^{\infty}_{0}{V_{\rm{GOP}}(R) R^{2} dR}/A_{p}A_{t}.
\end{equation}
The solid, dotted, and dashed curves are the results of GOPs for the $^{16}$O + $^{12}$C, $^{16}$O + $^{16}$O, and $^{12}$C + $^{12}$C systems, respectively.
The open circle, open square, and filled square are the results of POPs for the $^{16}$O + $^{12}$C, $^{16}$O + $^{16}$O, and $^{12}$C + $^{12}$C systems, respectively.
The ambiguity of the imaginary part is considered to be large as shown in Fig.~\ref{figXX} and mentioned in the previous paragraph.
Therefore, the volume integral per nucleon pair only for the real part is shown in Fig.~\ref{figYY}. 
The calculated volume integral per nucleon pair by GOP is consistent with the results by POPs.
In addition, it is found that the volume integral per nucleon pair smoothly shifts from attractive to repulsive in the same way as shown in Ref.~\cite{FUR10}.

\begin{figure}[tbh]
\begin{center}
\includegraphics[width=7cm]{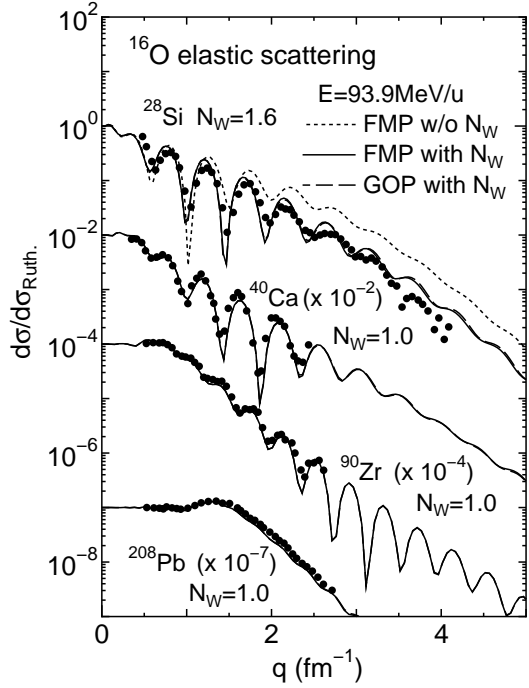}
\caption{\label{fig06} Elastic scattering of the incident $^{16}$O particle for various targets at $E$ = 93.9 MeV/u. 
The solid and dotted curves are the results by the FMP with the best fit and constant values ($N_{W}$ = 1.0), respectively. 
The dashed curves are the results by the present GOP with the $N_{W}$ value. 
The experimental data is taken from Ref.~\cite{ROU88}.}
\end{center} 
\end{figure}

\begin{figure}[tbh]
\begin{center}
\includegraphics[width=7cm]{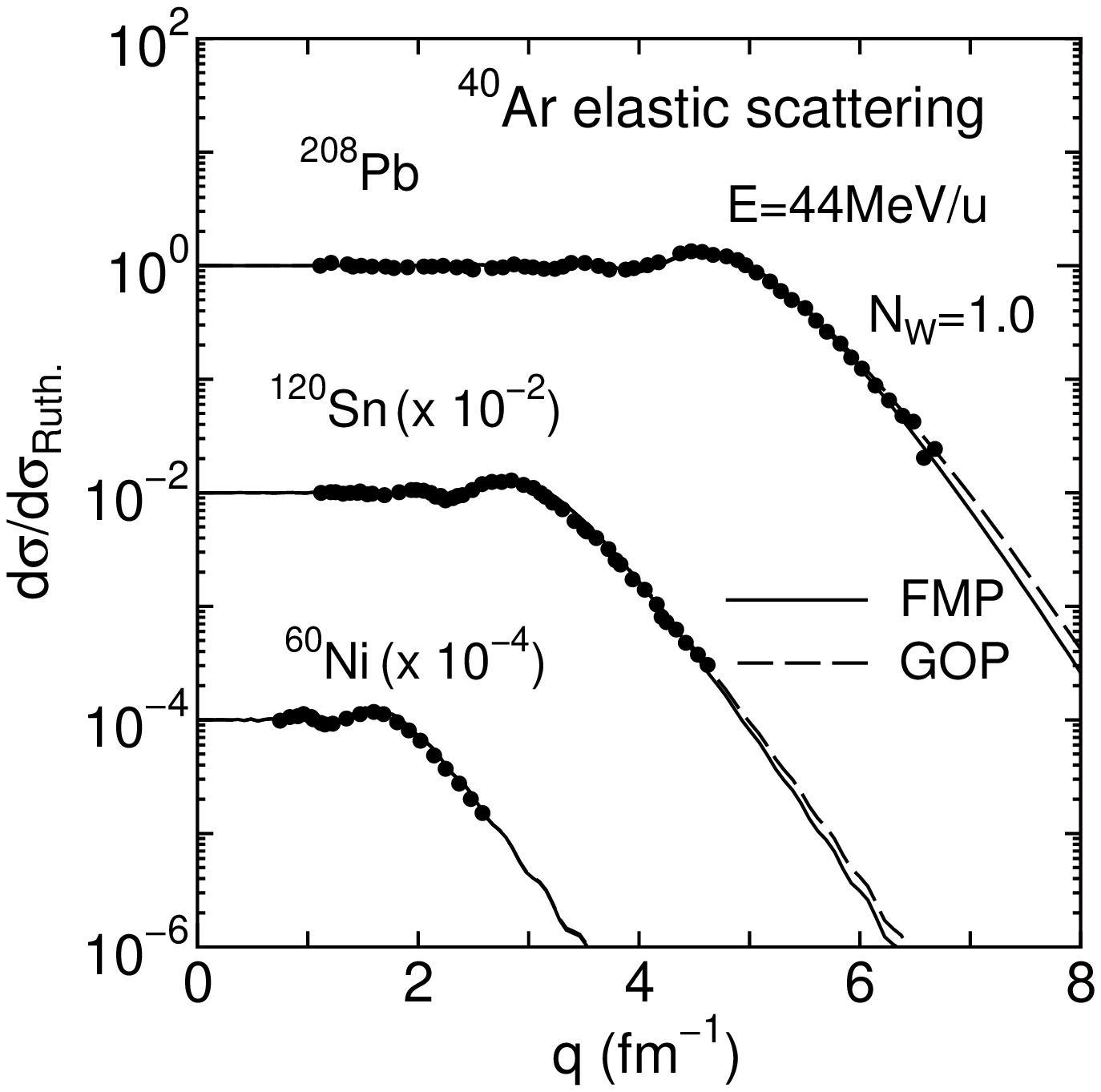}
\caption{\label{fig07} Same as Fig.~\ref{fig06} but for the $^{40}$Ar nucleus at $E =$ 44 MeV/u. 
The experimental data is taken from Ref.~\cite{ALA84}}
\end{center} 
\end{figure}

We also investigate the scattering by heavier target nuclei. 
Figures~\ref{fig06} and \ref{fig07} show the elastic scattering of $^{16}$O and $^{40}$Ar projectiles by various targets at $E$ = 93.9 and 44 MeV/u, respectively. 
The FMP and the present GOP well reproduce the data up to backward angle without $N_{W}$, except for the $^{16}$O scattering by $^{28}$Si case where a stronger absorption ($N_{W}=1.6$) is necessary to reproduce the data at backward angles. 
The origin of the exceptionally large value of $N_{W}$ for the $^{28}$Si target is unclear at present but the similar tendency was also reported in the case of $\alpha$ scattering by $^{28}$Si~\cite{FUR06} where the effect of strong collective excitation of the $^{28}$Si target was suggested.

\subsection{Unstable nuclear region}
In this section, we compare with the elastic cross sections calculated by the S{\~a}o Paulo and Niigata densities. 
With the S{\~a}o Paulo and Niigata densities, the elastic scatterings of the Carbon isotopes, Oxygen isotopes, and $^{30}$Ne nucleus by the double-magic nuclear targets ($^{16}$O, $^{40}$Ca, and $^{208}$Pb) at 100, 200, 300, and 400 MeV/u are compared with the results by the present GOP. 
We fix the $N_W$ value to be unity. 

\begin{figure}[tbh]
\begin{center}
\includegraphics[width=7cm]{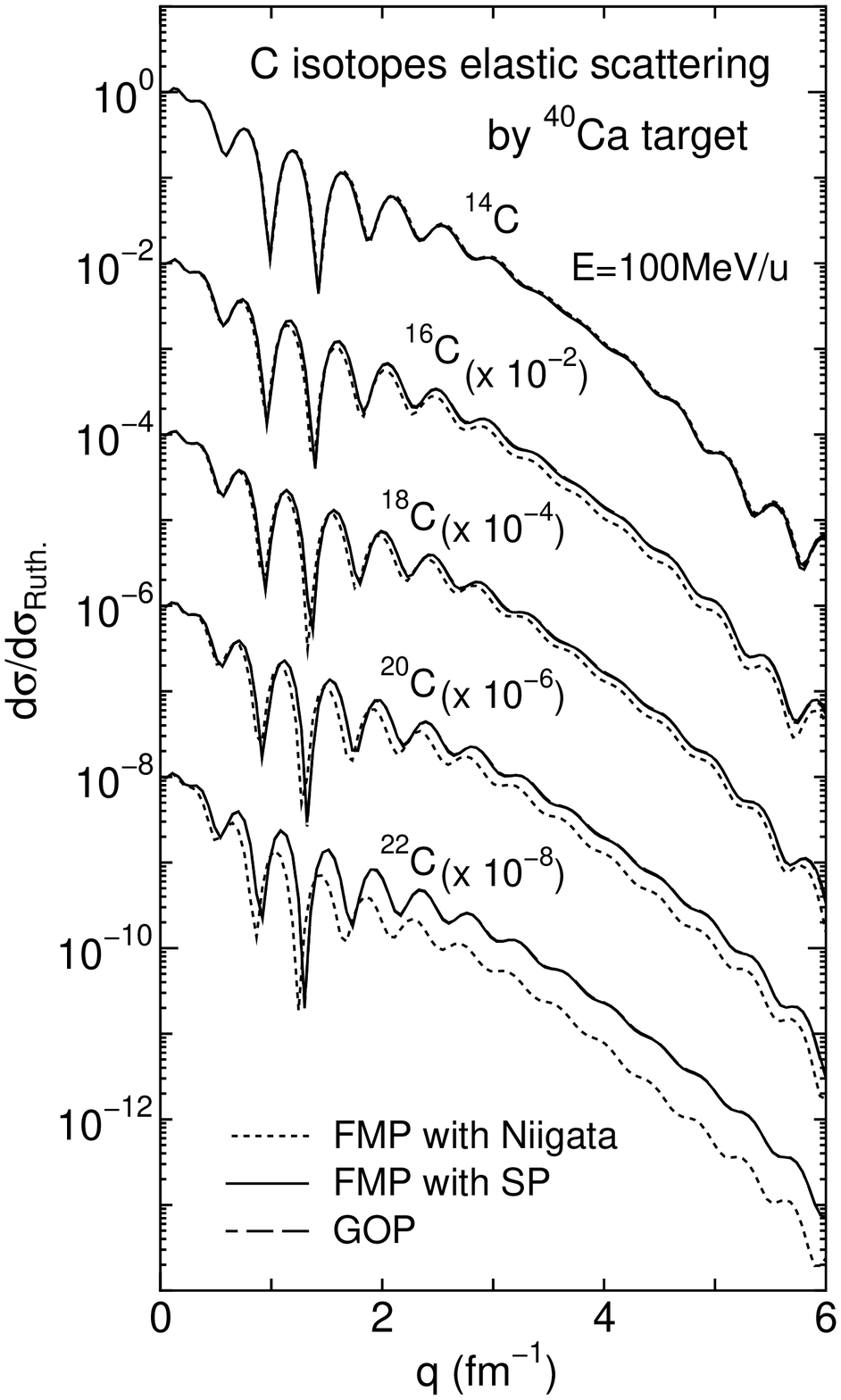}
\caption{\label{fig08} Comparison of the results by Niigata density with those by S{\~ a}o Paulo (SP) density in the elastic scattering cross sections of the incident C isotopes by $^{40}$Ca target at 100 MeV/u.}
\end{center} 
\end{figure}

\begin{figure}[tbh]
\begin{center}
\includegraphics[width=7cm]{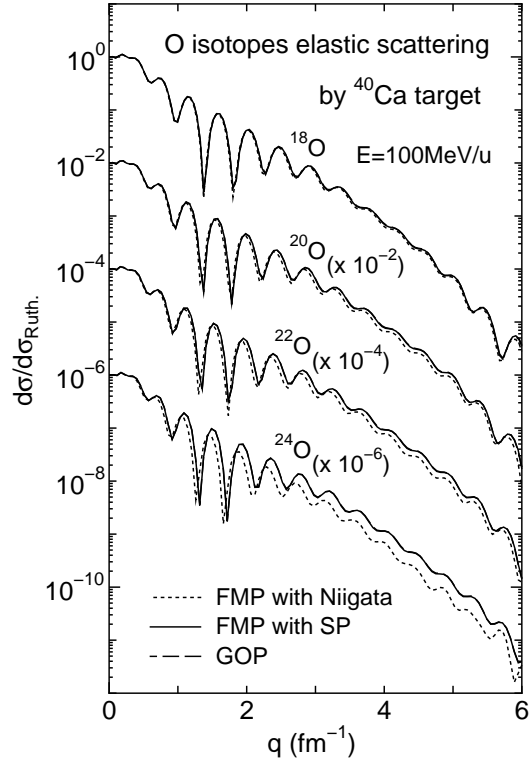}
\caption{\label{fig09} Same as Fig.~\ref{fig08}, but for the incident O isotopes.}
\end{center} 
\end{figure}

\begin{figure}[tbh]
\begin{center}
\includegraphics[width=7cm]{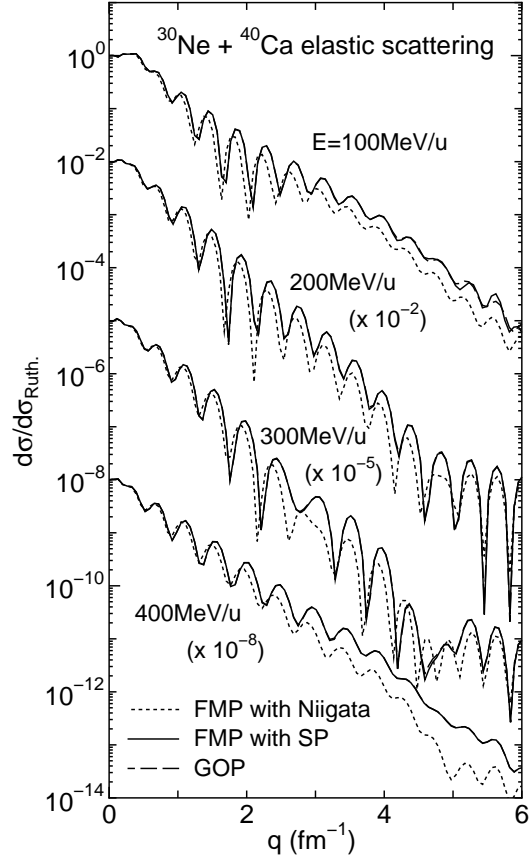}
\caption{\label{fig10} Comparison the result by Niigata density with that by S{\~ a}o Paulo (SP) density in elastic scattering cross section of the incident $^{30}$Ne particle by the $^{40}$Ca target at various energies.}
\end{center} 
\end{figure}

\begin{figure}[tbh]
\begin{center}
\includegraphics[width=7cm]{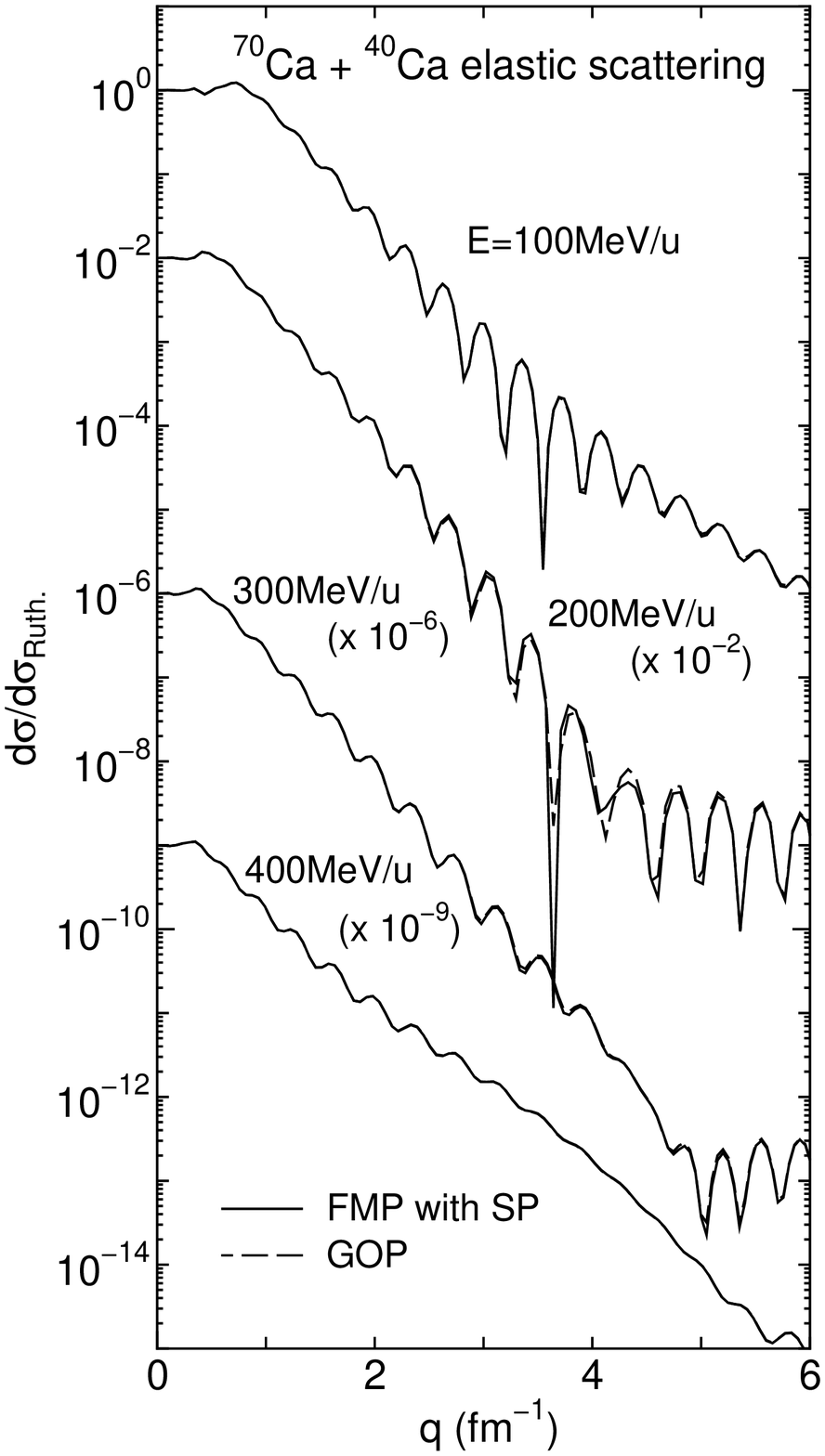}
\caption{\label{fig11} Same as Fig.~\ref{fig10} but for the $^{70}$Ca nucleus.}
\end{center} 
\end{figure}

Figure \ref{fig08} shows the elastic scattering of the incident $^{14-22}$C nuclei by the $^{40}$Ca target at 100 MeV/u. 
The dotted and solid curves are the results by the FMP with the Niigata and S{\~a}o Paulo densities, respectively. 
The dashed curves are the results by the present GOP. 
In this comparison,  the FMP and GOP well reproduce the elastic scattering cross section calculated with the Niigata density, except for the result of the incident $^{22}$C nucleus. 
The $^{22}$C nucleus is considered and observed to have a two neutron halo structure~\cite{HOR06-2,TAN10} and the S{\~a}o Paulo density can not describe such exotic structure, that is why these densities give different results in the elastic cross sections. 
For the incident $^{22}$C particle, the results of the GOP and FMP with S{\~a}o Paulo density overestimate that of the FMP with the Niigata density at backward angles. 
This reason comes from the difference of the diffuseness of the densities. 
In this paper, we present the calculated results only in the case by the $^{40}$Ca target at 100 MeV/u. 
The others which are not shown are available in Ref.~\cite{furumoto}. 

Figure \ref{fig09} shows the elastic scattering of the incident $^{18-24}$O nuclei by the $^{40}$Ca target at 100 MeV/u. 
For the Oxygen isotopes, the elastic cross section calculated with S{\~a}o Paulo density well reproduced the results with the Niigata one. 
In the case of the incident $^{24}$O particle, the difference is slightly seen in the cross section around backward angles. 
Because the neutron dripline nucleus of the Oxygen isotope ($^{24}$O) is not very weakly binding system (B.E. $\sim$ 4 MeV), that is, has no halo structure, the difference of the results between the S{\~a}o Paulo and Niigata densities is not as large as in the case of $^{22}$C. 

Figure~\ref{fig10} shows the elastic scattering of the incident $^{30}$Ne particle by the $^{40}$Ca target at various energies. 
Although the neutron number of the $^{30}$Ne nucleus is a magic number, 
the shell gap vanishes around the $^{30}$Ne nucleus and the region is known to be island of inversion~\cite{DOO09}. 
Therefore, the relation between the elastic cross sections with SP
density and those with Niigata density of $^{30}$Ne is similar to the $^{24}$O case.
At 400 MeV/u, the result by S{\~a}o Paulo density reproduces that by Niigata
density for the forward angles ($q < 3$), but the different results are
given for the backward angles ($q > 3$). 
A similar result is obtained in the case of the incident $^{24}$O particle at 400 MeV/u~\cite{furumoto}. 

Finally, Fig.~\ref{fig11} shows the elastic scattering of $^{70}$Ca by the $^{40}$Ca target at various energies. 
According to Refs.~\cite{MOL95,MOL97,KOU00,KOU05}, we assume that the $^{70}$Ca nucleus is set to be a dripline nucleus.
Moreover, this nucleus is the heaviest projectile in this paper. 
For the heaviest projectile, the GOP well reproduce the elastic cross section calculated with FMP. 
Namely, it imply that the fitting accuracy of GOP is confirmed in the wide range region of the projectile.

\section{conclusion}
We have constructed a new global optical potential (GOP) in the framework of the CEG07 folding model with S{\~ a}o Paulo density. 
The incident particles are even-even isotopes of $^{8-22}$C, $^{12-24}$O, $^{16-38}$Ne, $^{20-40}$Mg, $^{22-48}$Si, $^{26-52}$S, $^{30-62}$Ar, and $^{34-70}$Ca.
We set the reliable energy range of GOP to be 50 -- 400 MeV/u from the comparison with available experimental data.
The present GOP can be used for heavy ion projectiles of stable and unstable nuclei over the wide range of nuclear chart with $Z=6\sim 20$, except for 
those on or near the driplines.
The GOP parameters and the program source for constructing the GOP from the parameters are available in Ref.~\cite{furumoto}.

\section{Acknowledgment}
The authors would like to thank to Doctor Takatoshi Ichikawa for useful advices and comments. 
One of the present authors (W.H.) is supported by the Special Postdoctoral Researchers Program of RIKEN.
The numerical computation in this work was carried out at the Computer Facilities at Yukawa Institute for Theoretical Physics and Nuclear Theory Group of Osaka City University.


\end{document}